\documentstyle[12pt]{article}
\begin{document}
\def\singlespacing{\baselineskip=12pt}
\def\doublespacing{\baselineskip=24pt}
\def\f#1{{\tilde{\mbox{\boldmath $#1$}}}}
\def\i{{\mbox{\boldmath $i$}}}
\def\d{{\tilde{\mbox{\boldmath $d$}}}}
\def\real{\hbox{\rlap{\vrule height6.7pt width0.4pt}R}}
\def\dx{\d{x}}
\def\dt{\d{t}}
\hyphenation{mo-no-poles}
\def\e{{\tilde{\mbox{\boldmath $e$}}}}
\def\E{{}^{(4)}\e}
\singlespacing
\noindent
\pagestyle{empty}
{}\hfill\\
\def\K{{\bf K}}
\def\inv{\f\omega}
\def\O{{\cal O}}
\def\Ls{\pounds}
\bigskip
\bigskip
\doublespacing
\font\mainh=cmbx10 scaled \magstep1
\baselineskip=24.9truept plus 2pt
\def\J{{\bf J}}
\def\frac#1#2{{\textstyle {#1\over #2}}}
\medskip
\begin{center}
\begin{large}
{\mainh The Hartle-Hawking State for the Bianchi IX Model in Supergravity}\\
\end{large}
\bigskip
\bigskip
R. GRAHAM\\
\smallskip
Fachbereich Physik\\
Universit\"at Gesamthochschule Essen\\
D-45117 Essen \\
Germany\\
\bigskip
\medskip
H. LUCKOCK\\
\smallskip
School of Mathematics and Statistics\\
University of Sydney\\
NSW 2006\\
Australia\\
\end{center}
\bigskip
\bigskip
\bigskip
\leftline{\bf Abstract}\smallskip\noindent

We solve the quantum constraints for homogeneous N=1 supergravity
on  3-geometries with a Bianchi IX metric. Because these geometries
admit Killing vectors with the same commutation relations as the
angular momentum generators, there are two distinct definitions
of homogeneity. The first of these is well-known and has been shown
by D'Eath to give the wormhole state. We show that the alternative
definition of homogeneity leads to the Hartle-Hawking ``no boundary"
state.\hfill

\bigskip
\leftline{PACS numbers: 98.80.H , 04.65}

\newpage
\pagestyle{plain}

\pagenumbering{arabic}

\noindent
Minisuperspace models have attracted interest for a long time as a
useful testing ground for new ideas on the many unresolved questions
in quantum gravity \cite{DeWitt,Misner,HartleHawking}.  Recently a
variety of minisuperspace models have been shown to admit simple
supersymmetric extensions \cite{MaciasObregonRyan,D'EathHughes,
Gr1,Gr4,GrBene,D'EathHawkingObregon} and it appears that many of
these can be derived from a full supergravity theory by dimensional
reduction of the classical theory \cite{MaciasObregonRyan,D'EathHughes}.

A slightly different approach is to start with the supersymmetry
constraints for full supergravity and then to consider only
homogeneous configurations of the metric and the fields. This
approach has been used recently by D'Eath to quantise Bianchi IX
cosmologies \cite{D'Eath2} (see also \cite{Asano}). D'Eath showed
that if the components of the Rarita-Schwinger field in an invariant
basis are required to be spatially constant, then the supersymmetry
constraints can be solved by a wormhole state but, surprisingly, not
by a Hartle-Hawking state.

This result is puzzling, as it calls into question the existence of a
Hartle-Hawking state for homogeneous $N=1$ supergravity. The aim of
this letter is to solve this puzzle. We will show that the existence
of a Hartle-Hawking state depends on the ansatz used for the fermion
fields; specifically, on whether the spinor components have the same
or opposite sign at antipodal points of the spatial 3-manifold.

The spatial 3-geometry in Bianchi IX models is homeomorphic either to
$SO(3)$ or to its double covering $SU(2)\sim {\cal S}^3$. The second
choice is necessary if we seek a homogeneous Hartle-Hawking state
since $SU(2)$ (unlike $SO(3)$) can be continuously shrunk to a point
within some smooth 4-manifold.

Identifying points in ${\cal S}^3$ with matrices $U\in SU(2)$, the
symmetry transformations for homogeneous geometries can be
represented by mappings $U\mapsto UB^\dagger$ where $B\in SU(2)$.  A
field $\phi$ can be said to be homogeneous if it is invariant, up to
the action of some locally-acting group, under these transformations;
in other words, if it satisfies
$$\phi(UB^\dagger) = R_B \phi(U)$$
where $R_B$ is some representation of the matrix $B\in SU(2)$.

The most obvious definition of homogeneity arises from the trivial
representation, in which $R_B=1$ for all $B\in SU(2)$; this choice
leads to D'Eath's ansatz, in which all fields have constant
components. However another choice is possible, since physical fields
carry a representation of the group of spatial rotations which are
themselves represented by elements of $SU(2)$. An alternative
definition of homogeneity can thus be obtained by choosing $R_B$
to be the rotation represented by the matrix $B$ (or by matrices
obtained from $B$ by unitary transformations).

The distinction between the two kinds of homogeneity is clearest in
the case of a spin-${1\over 2}$ field $\chi$ on ${\cal S}^3$. Instead
of requiring $\chi$ to have constant components when referred to an
invariant dreibein (i.e.  a dreibein invariant under the
diffeomorphisms $U\mapsto UB^\dagger$), we can demand that $U\chi$
should be constant on ${\cal S}^3$.

For our purposes, it will be useful to obtain a differential version
of the new homogeneity condition. The spatial 3-geometry of the
Bianchi IX model admits three Killing vector fields $\K_p$ and three
invariant 1-forms $\inv^p$ satisfying
$$[\K_p,\K_q]= {C^r}_{pq} \K_r,\ \ \ \
\d\inv^p= {1\over 2} C^p{}_{qr} \inv^q\wedge\inv^r\ \ \ \
{C^r}_{pq} =\epsilon_{pqs} \delta^{sr}$$
while the 3-metric has the form $h=h_{pq}\inv^p\otimes\inv^q$.
The components of $h$ in the invariant basis can be written as
$$h_{pq}= e^{2\beta_a} \O_p{}^a\O_q{}^a$$
where the diagonalising matrix $\O_q{}^a$ satisfies the orthogonality
condition
$$\delta ^{pq} \O_p{}^a \O_q{}^b = \delta^{ab},\ \ \ \ \ \ \det\O = 1.$$
Apart from inversions and relabellings of the axes, there is a unique
matrix $\O$ and a unique set of scale parameters $\beta_a$ corresponding
to a given 3-metric $h$.

The three scale parameters $\beta_a$ represent the physical degrees of
freedom of the 3-metric. On the other hand, the three degrees of freedom
contained in the matrix $\O$ are pure gauge, and are associated with
the group of diffeomorphisms generated by the three Killing vector
fields dual to the 1-forms $\inv^p$. By taking the matrix $\O$ to be
fixed, we eliminate these unphysical degrees of freedom.

We can now identify a preferred Lorentz frame in which the axes of
the dreibein coincide with the main axes of the metric tensor; in this
special gauge the dreibein consists of the three 1-forms
\begin{equation}
\e^a = e^a{}_p\inv^p,\ \ \ \ \ \ \ \ e^a{}_p= e^{\beta_a}\O_p{}^a.
\label{eq:dreibein}
\end{equation}
We will work exclusively in this preferred frame, thereby eliminating
the gauge degrees of freedom associated with Lorentz transformations.

In the classical theory, the 3-geometry is embedded in a
4-dimensional spacetime equipped with a vielbein consisting of
four 1-forms which we take as
$$\E^0= N\dt ,\ \ \ \ \ \ \
\E^a= e^a{}_p(\inv^p + N^p \dt)\ \ \ \ (a=1,2,3).$$
We are entitled to impose a coordinate condition on the lapse
function $N$, in addition to the three we have already imposed
by fixing the matrix $\O$. Together with the three scale parameters
$\beta_a$, the components $N^p$ of the shift vector then account for
the six physical degrees of freedom of the 4-metric. Note however that
the wave function is independent of $N$ and the variables $N^p$, and so
these do not appear in the quantum theory which we consider here. We
are therefore interested only in the intrinsic geometry of the spatial
3-manifold, which is fully described by the dreibein $\e^a$.

In the preferred Lorentz frame, the new homogeneity condition for an
arbitrary field $\phi$ defined on the spatial ${\cal S}^3$ manifold
has the form
\begin{equation}
\Ls_{\K_p} \phi =- i (i_{\K_p}\inv^q) \O_q{}^a \J^a \phi \ \ \ \ \ \ \ \ \ \ \
p=1,2,3
\label{eq:hom}
\end{equation}
where $i_{\K_p} \inv^q$ denotes the interior product of $\K_p$ with
$\inv^q$, and $\Ls_{\K_p}$ is the Lie derivative along $\K_p$. (The
dreibein is invariant under the action of the $\K_p$ and so
$\Ls_{\K_p}$ does not contain a spin connection\cite{BennTucker}.)
The $\J^a$ generate physical rotations of $\phi$ about the axes defined
by the 1-forms $\e^a$ and obey the standard commutation relations
$$[\J^a,\J^b] =i\epsilon^{abc} \J^c.$$
These relations and the orthogonality of the matrix $\O$ ensure that
the new homogeneity condition is integrable and admits non-trivial
solutions.

Condition (\ref{eq:hom}) implies that $\phi$ rotates through an angle
$2\pi$ relative to an invariant frame as one follows a path between
antipodal points on ${\cal S}^3$. In particular, this means that the
components of any homogeneous spinor field will have opposite signs
at antipodal points.

Since the operators appearing on both sides of (\ref{eq:hom}) obey
the Leibnitz rule, it is clear that products of homogeneous fields
will themselves be homogeneous. The 3-metric $h$ is also homogeneous
since $\Ls_{\K_p}h=0=\J_qh$.

Of particular interest in supergravity is the spatial part
$\psi^A=\psi^{Aa}\e^a$ of the spin-${3\over 2}$ Rarita-Schwinger field
which appears in the quantum supersymmetry constraints.
Introducing local coordinates $x^i$ on the 3-manifold, with
$\inv^p=\omega^p{}_i \dx^i$, the homogeneity conditions for
the components $\psi^{Aa}$ can be written as
\begin{equation}
{\partial \psi^{Aa}\over\partial x^i} = - i \omega^p{}_i \O_p{}^b
\left(  {1\over \sqrt 2}\sigma_b{}^{AA'} \delta _{BA'}\psi^{B\,a} +
i\epsilon^{abc}
 \psi^{Ac}\right).
\label{eq:hom_spin}
\end{equation}
We use here the two-component spinor notation\cite{D'Eath1}, in which
the Infeld-van der Waerden symbols $\sigma_b{}^{AA'}$ represent the components
of the matrices ${1\over\sqrt 2} \tau^b$.

We now determine the form of the quantum supersymmetry generators for
$N=1$ supergravity subject to these conditions. Rewriting the fields
in terms of the local coordinate basis as $\psi^A=\psi^A{}_i\dx^i$ and
$\e^a= e^a{}_i\dx^i$, and defining $e^{AA'}{}_i \equiv \sigma_a{}^{AA'}
e^a{}_i$,
these generators have the form\cite{D'Eath1,quanrul}
$$\overline S_{A'} = \epsilon^{ijk}\, e_{AA'\,i}\, D_j  \psi^A{}_k\
+{\kappa^2\over 2}  \psi^A_i\, {\delta \over \delta e^{AA'}{}_i} $$
and
$$S_A =D_i {\delta \over \delta \psi^A{}_i } +
{\kappa^2\over 2} {\delta \over \delta e^{AA'}{}_i}
\left( D^{BA'}{}_{ji} {\delta\over\delta\psi^B{}_j}\right)$$
where $\kappa^2= 8\pi$ and $D_i$ denotes the torsion-free spatial covariant
derivative. An expression for the kernel $D^{AA'}{}_{ji}$ is given
in \cite{D'Eath1}.

Supersymmetry imposes on the wave function $\Psi(e^{AA'}{}_i,\psi^A{}_i)$
the constraints $$\overline S_{A'}\Psi = 0 =S_A \Psi .$$ Together
with the angular momentum constraints, these ensure that all the
remaining constraints of the theory are also satisfied. For the
mini-superspace model discussed here,  it will be enough
to consider the zero-fermion state, which is annihilated
by the operators ${\delta\over\delta\psi^A{}_i}$ and whose  wave
function $\Psi_0$ therefore depends only on the 3-geometry.

For this state, the $S_A$ constraint is satisfied automatically.
Following D'Eath \cite{D'Eath2}, we evaluate the other supersymmetry
constraint subject to the homogeneity conditions. After some
algebra, the $\overline S_{A'}$ constraint may be rewritten as
$$- \det
[\omega^p{}_i] \left(  {1\over 2} e_{AA'\, s}\delta^{sr} \psi^A{}_r
+{i\lambda \over\sqrt 2}\epsilon^{pqr} e_{BA'\,p} \O_q{}^a\sigma_a{}^{BB'}
\delta_{AB'}\psi^A{}_r
\right)\Psi_0 ={\kappa^2\over 2} \psi^A{}_r { \delta \Psi_0 \over
\delta e^{AA'}{}_r}$$
where $e^{AA'}{}_p \omega^p{}_i= e^{AA'}{}_i$
and $\psi^A{}_p\omega^p{}_i= \psi^A{}_i$.
Here we have introduced a parameter $\lambda$ which is 1 for our
ansatz (\ref{eq:hom}), and which vanishes for the ansatz used by D'Eath.
We note that both terms in the bracket of eq.~(\ref{eq:hom_spin})
have been combined in the single term proportional to $\lambda$.

Cancelling the factors of $\psi^A{}_r$ which appear on both sides, and
integrating over the whole of ${\cal S}^3$ we obtain $${\kappa^2\over
2} {\partial (\ln |\Psi_0|) \over \partial e^{AA'}_r} = (4\pi)^2
\left(  {1\over 2} e_{AA'\, s}\delta^{rs}
+ {i\lambda\over \sqrt 2}
\epsilon^{pqr}e_{BA'\,p} \O_q{}^a \sigma_a{}^{BB'} \delta_{AB'}\right).$$
If the dreibein has the form given in (\ref{eq:dreibein}) then this
expression can be integrated to give
\begin{equation}
\Psi_0=  const \times \exp \left[  -\pi\left(e^{2\beta_1} + e^{2\beta_2}
+e^{2\beta_3} -2\lambda e^{\beta_1+\beta_2} -  2\lambda
e^{\beta_2+\beta_3}- 2\lambda e^{\beta_3+\beta_1} \right) \right].
\label{eq:soln}
\end{equation}
For $\lambda=0$ this is the solution first given in \cite{Ryan} and
\cite{Gr1}, and later derived in \cite{D'Eath2} and \cite{Asano} by
the same technique as used here.  The $\lambda=0$ solution is now
commonly referred to as the wormhole state, as its exponent defines a
Euclidean Hamilton-Jacobi action which gives rise to Riemannian
4-geometries which are asymptotically flat but become singular as the
scale parameter $\exp (\beta_1+\beta_2+\beta_3)$ approaches zero
\cite{GibbonsPope,Belinsky}. The latter property forbids the
interpretation of this solution as a Hartle-Hawking state, as noted
by D'Eath \cite{D'Eath2}.

On the other hand, our new homogeneity condition corresponds to the
choice $\lambda=1$.  In this case our solution (\ref{eq:soln}) has
all the features of a Hartle-Hawking state. (This solution
has been previously obtained from an $N=2$ supersymmetric extension
of the bosonic Wheeler DeWitt equation for Bianchi IX minisuperspace
\cite{GrBene}.) In particular, when restricted to the isotropic case
$\beta_1= \beta_2= \beta_3=\alpha$, $\Psi_0$ grows with $\alpha$ like
$\exp(3\pi e^{2\alpha})$ which is a well-known feature of the
Hartle-Hawking state in isotropic Bianchi IX models
\cite{HartleHawking}. Furthermore, by using the exponent of (\ref{eq:soln})
as a Euclidean Hamilton-Jacobi action, one generates a 1-parameter
family of 3-metrics on ${\cal S}^3$ which remain regular for small
scale parameters \cite{GibbonsPope, AtiyahHitchin}.

We note also that both states $\lambda=0,1$ are normalisable for fixed
$\alpha$ when integrated over $\beta_+={1\over 6}
(\beta_1+\beta_2-2\beta_3)$ and $\beta_-= (\beta_1-\beta_2)/2\sqrt{3}$,
thus allowing a straightforward probabilistic interpretation.

Let us finally remark on solutions in the other fermion sectors. As
was first shown for the Bianchi I model in
\cite{D'EathHawkingObregon} and later for Bianchi IX models with
$\lambda=0$ in \cite{D'Eath2,Gr4}, the constraint of Lorentz
invariance, automatically satisfied by $\Psi_0$ which depends only on
the 3-geometry, rules out states in any but the empty and filled
fermion sectors. The same must happen in the $\lambda=1$ case
considered here.

Turning to the wave function $\Psi_f$ in the filled sector, we can
use the duality between particles and holes to conclude that $\Psi_f
\sim \Psi_0^{-1}$.  Because the states $\Psi_0$ are normalisable for
fixed $\alpha$, the states $\Psi_f$ are not; they diverge for
infinite anisotropies and must therefore be disregarded\cite{Gr1}.

In conclusion, we have shown that imposing the supersymmetry
constraints for $N=1$ supergravity on Bianchi IX 3-geometries and
spinors with the homogeneity property (\ref{eq:hom}) yields a unique
quantum state with the properties specified by the no-boundary
proposal of Hartle and Hawking.
\bigskip
\bigskip
\bigbreak

\leftline{\bf Acknowledgements}
This work was supported by a Sydney University Research Grant, and by
the Deutsche Forschungsgemeinschaft through the
Sonderforschungsbereich 237 ``Unordnung und Grosse Fluktuationen".
R.G. wishes to thank the School of Mathematics at the University of
Sydney for hospitality, and to acknowledge useful discussions with
Peter D'Eath.

\vfill

\begin{thebibliography}{20}
\bibitem{quanrul} Here we correct a sign error in the second term on
the right-hand side of eq. (4.6) of \cite{D'Eath1} which was taken
over in \cite{D'EathHawkingObregon,Asano}. In order not to obscure
the comparison with refs. \cite{D'EathHawkingObregon,Asano} we shall
compensate for this change of sign by choosing $\det[\omega^p{}_i]<0$.
\end{thebibliography}
\end{document}